\documentclass{aa-package/aa}

\usepackage{color}
\usepackage{colortbl}
\usepackage{multirow}
\usepackage{dcolumn}
\usepackage{amsmath}
\usepackage{amssymb}
\usepackage{graphicx}
\usepackage{epsfig}
\usepackage{changebar}
\usepackage{natbib}
\usepackage{multirow}
\usepackage{subfigure}
\usepackage{txfonts}
\usepackage{relsize}

\usepackage{aa-package/aalongtable,lscape}
\usepackage{longtable,lscape}
\usepackage[figuresright]{rotating}

\newcolumntype{d}[1]{D{.}{\cdot}{#1}}

\newcolumntype{.}{D{.}{.}{-1}}

\newcommand{\lsun}{L$_\odot$}

\newcommand{\vlsr}{V$_{\rm{LSR}}$}

\newcommand{\mum}{$\mu$m}

\newcommand{\kms}{km~s$^{-1}$}

\begin{document}
   \title{The RMS Survey}

  \subtitle{H$_2$O masers towards a sample of southern hemisphere massive YSO candidates and ultra compact HII regions\thanks{Table 1 and 2 and full version of Fig.~2 are only available in electronic form at the CDS via anonymous ftp to cdsarc.u-strasbg.fr (130.79.125.5) or via http://cdsweb.u-strasbg.fr/cgi-bin/qcat?J/A+A/.}}

   \author{J.~S.~Urquhart
          \inst{1,2}  
			 \and
			 M.~G.~Hoare
			 \inst{1}
			 \and
			 S.~L.~Lumsden
			 \inst{1}
			 \and
			 R.~D.~Oudmaijer
			 \inst{1}
			 \and
			 T.~J.~T.~Moore
			 \inst{3}
			 \and
			 P.~R.~Brook
			 \inst{1} 
			 \and
			 J.~C.~Mottram
			 \inst{1,4} 
			 \and
			 B.~Davies
			 \inst{1} 
			 \and
			 J.~J.~Stead
			 \inst{1} 
          }

     \offprints{J. S. Urquhart: james.urquhart@csiro.au}

   \institute{School of Physics and Astronomy, University of Leeds, Leeds, LS2~9JT, UK 
         \and Australia Telescope National Facility, CSIRO, Sydney, NSW 2052, Australia   
         \and Astrophysics Research Institute, Liverpool John Moores University, Twelve Quays House, Egerton Wharf, Birkenhead, CH41~1LD, UK
		 \and School of Physics, University of Exeter, Exeter, EX7 4QL, UK  
             }

   \date{}

\abstract
   {The Red MSX Source (RMS) survey has identified a large sample of candidate massive young stellar objects (MYSOs) and ultra compact (UC) HII regions from a sample of $\sim$2000 MSX and 2MASS colour selected sources.}      
   {To search for H$_2$O masers towards a large sample of young high mass stars and to investigate the statistical correlation of H$_2$O masers with the earliest stages of massive star formation.} 
   {We have used the Mopra Radio telescope to make position-switched observations towards $\sim$500 UCHII regions and MYSOs candidates identified from the RMS survey and located between 190\degr $< l < $ 30\degr. These observations have a 4$\sigma$ sensitivity of $\sim$1~Jy and a velocity resolution of $\sim$0.4~\kms.}
   {We have detected 163 H$_2$O masers, approximately 75\% of which were previously unknown. Comparing the maser velocities with the velocities of the RMS sources, determined from $^{13}$CO observations, we have identified 135 RMS-H$_2$O maser associations, which corresponds to a detection rate of $\sim$27\%. Taking into account the differences in sensitivity and source selection we find our detection rate is in general agreement with previously reported surveys.}
   {We find similar detection rates for UCHII regions and MYSOs candidates, 
suggesting that the conditions needed for maser activity are equally likely in these two stages of the star formation process. Looking at the detection rate as a function of distance from the Galactic centre we find it significantly enhanced within the solar circle, peaking at $\sim$37\% between 6--7~kpc,  which is consistent with previous surveys of UC HII regions, possibly indicating the presence of a high proportion of more luminous YSOs and HII regions.}

\authorrunning{J. S. Urquhart et al.}
\titlerunning{H$_2$O masers towards young massive stars}
\maketitle

\section{Introduction}

Massive young stellar objects (hereafter MYSOs) are in the early phase in the life of OB stars when fusion has most likely started in the core, but when they have not yet begun to ionize their surroundings to form an HII region. The embedded mid-infrared point source usually possesses a strong ionized stellar wind (e.g., \citealt{bunn1995}) and drives bipolar molecular outflows (e.g., \citealt{lada1985,bronfman2008}; and reference therein). This brief (10$^{4-5}$ years) phase is clearly crucial to our understanding of how these massive young stars form since during this time any ongoing accretion will be halted and the final mass of the star is set. In addition, the winds, outflows and eventual HII regions have an important feedback role in determining the fate of the rest of the molecular cloud. 

The main difficulty in studying MYSOs is their relative rarity. Most current work still relies on samples drawn from the IRAS Point Source Catalogue, but whose selection criteria bias them away from complex regions by requiring them to be away from HII regions identified in single-dish radio surveys (e.g., \citealt{molinari1996, sridharan2002}). Additionally, the known samples are too small to test many aspects of massive star formation theories, and are probably unrepresentative of the class as a whole. In an effort to address these issues we have conducted a new galaxy-wide search for MYSOs starting from the MSX mid-infrared survey of the Galactic Plane (\citealt{price2001}). This has significantly better spatial resolution than IRAS and so does not have the same confusion problems. We have used the MSX Point Source Catalogue (\citealt{egan2003}) to colour-select a large sample of candidate MYSOs (\citealt{lumsden2002}). This initial sample was further refined using 2MASS data to eliminate blue objects and from visually inspecting the MSX images to remove move extended sources. Our colour-selection and subsequent filtering produced a sample of $\sim$2000 MYSO candidates (\citealt{lumsden2002}).

The Red MSX Source (RMS) Survey is a multi-wavelength programme of follow-up observations designed to identify genuine MYSOs and ultra compact (UC) HII regions from the other kinds of embedded or dusty objects such as planetary nebulae (PNe), evolved stars and nearby low-mass YSOs (\citealt{hoare2005,mottram2006,urquhart2007c}). The main aim of the project is to produce a large sample of MYSOs and UCHII regions, and to compile a database of complementary multi-wavelength data with which to study their properties.\footnote{http://www.ast.leeds.ac.uk/RMS} We have used arcsecond resolution mid-infrared imaging from the Spitzer GLIMPSE survey (\citealt{benjamin2003}) or our own ground-based imaging (e.g., \citealt{mottram2007}) to reveal multiple and/or extended sources within the MSX beam, as well as MYSOs in close proximity to existing HII regions. We have obtained arcsecond resolution radio continuum data with ATCA and the VLA (\citealt{urquhart_radio_south, urquhart_radio_north}) to identify UCHII regions and PNe, whilst observations of  $^{13}$CO transitions (\citealt{urquhart_13co_south, urquhart_13co_north}) deliver kinematic velocities. Finally we have obtained near-infrared spectroscopy (e.g., \citealt{clarke2006}) which allows us to identify the more pathological evolved stars.

The kinematic velocities can be used in conjunction with a Galactic   rotation model  (e.g., \citealt{brand1993,alvarez1990,clemens1985}) to derive kinematic distances and luminosities, which allow us to distinguish between nearby low- and intermediate-mass YSOs and genuine MYSOs. However, the velocity of sources located within the solar circle -- which accounts for $\sim$80\% of the sample -- results in two possible kinematic distances equally spaced on either side of the tangent position; these are referred to as the near and far distances. This distance ambiguity needs to be solved before luminosities can be calculated and our sample of MYSO candidates can be turned into a sample of bona fide MYSOs. We have solved the distance ambiguities towards a sample of MYSOs using the HI self-absorption (see \citealt{busfield2006} for details and a description of the technique) but distances to the whole sample are not currently available.

By combining these observational data sets it is possible to identify the different source populations and remove the contaminates. With the classification effectively complete we are now moving into the exploitation phase of the RMS survey. The first step is to examine the global characteristics of this galaxy-wide sample of massive young stars as a prelude to detailed studies of sub-samples. Part of this involves determining the physical and chemical nature of the environment as a way of gauging their evolutionary status. We have used the Mopra telescope and the large bandwidth available with the broadband ($\sim$8~GHz) MOPS spectrometer to survey a sub-sample of MYSO candidates and UCHII regions at 12~mm. The primary focus is to study water maser emission and hot ammonia, but also class~II methanol maser emission and other serendipitous molecular transitions found in the range. 

In this paper we present the results of our search for water masers made towards $\sim$500 MYSO candidates and UCHII regions observable by Mopra. Water masers are the most likely masing species to be found associated with the mid-infrared bright MYSO phase. VLBI proper motion measurements show that they arise in the main from the base of bipolar outflows and jets (\citealt{goddi2005,moscadelli2005}) and as such trace an important part of the physics of massive star formation. Although the spatial resolution is not sufficient to warrant a detailed study of individual sources, these single-dish observations allow us to look at the statistical occurrence of water masers towards massive star forming regions, as well as paving the way for future interferometric follow-up observations. 

\section{Observations and source selection}

\subsection{Source selection}
\label{sect:source_selection}

To date our ongoing programme of classification has led to the identification of  approximately 500 MYSO candidates and a further 600 UC HII regions. In addition to these categories we have two further classifications that are relevant to this study; these are the HII/YSO and Young/old classifications. The first of these is reserved for RMS sources where both an HII region and YSOs are found within the MSX beam (i.e., $<$18\arcsec). We have identified $\sim$200 of these, which  are interesting in that they may indicate sequential star formation. The second of these classes contains sources that appear
isolated in near-infrared images and which would normally indicate an
evolved star classification, however, they also display strong CO towards
them and/or lie on dark filaments seen at 8~$\mu$m. They could be luminous evolved background stars seen through a molecular cloud or be genuine YSOs forming in the cloud.  These cannot be unambiguously classified until infrared spectra have been obtained and so remain within this class until a reliable classification can be attributed.

Approximately 900 RMS sources classified as one of the four types mentioned in the previous paragraph are observable by the Mopra telescope (i.e., 190\degr\ $< l <$ 30\degr). However, due to a combination of limited observing time and density of sources towards the Galactic centre we were only able to observe a little over half this number. Since the overall aim of the RMS project is to identify a large sample of massive YSOs, priority was given to sources identified as either YSOs or HII/YSOs with $\sim$80\% and $\sim$90\% of these types observed respectively. A high priority was also given to the Young/old class of sources, with $\sim$80\% being observed, as a significant  number of genuine YSOs may still be included in this catagory. Many of the sources excluded from our observations of these three classes of objects were on the grounds that water masers have previously been reported in the literature, or because they are located within a region of the Galactic Plane which is already the focus of a blind 19--27~GHz survey currently underway (HOPS; \citealt{walsh2008}). HII regions were considered a lower priority and were observed when time and RA range became available; only approximately 30\% of the available HII regions have been observed.

\begin{table}
\begin{center}
\caption{Summary of sources observed, positions and sensitivities.}
\label{tbl:source_positions}
\begin{minipage}{\linewidth}
\begin{tabular}{cccc}
\hline
\hline
MSX name	& RA& Dec	& r.m.s \\
& J2000& J2000&  (mK) \\
\hline

G018.8330$-$00.3004	&	18:26:23.64	&	$-$12:39:40.3	&		23	\\
G019.7540$-$00.1279	&	18:27:31.48	&	$-$11:45:56.8	&		25	\\
G027.9334+00.2056	&	18:41:34.15	&	$-$04:21:11.8	&		14	\\
G289.1447$-$00.3454	&	10:57:07.24	&	$-$60:07:15.5	&		26	\\
G293.5607$-$00.6703	&	11:30:07.10	&	$-$62:03:12.9	&		24	\\
G296.8926$-$01.3050	&	11:56:49.89	&	$-$63:32:04.9	&		24	\\
G300.7221+01.2007	&	12:32:50.83	&	$-$61:35:28.6	&		23	\\
G301.8147+00.7808	&	12:41:53.78	&	$-$62:04:13.0	&		26	\\
G305.9402$-$00.1634	&	13:17:52.94	&	$-$62:52:50.5	&		25	\\
G311.5131$-$00.4532	&	14:05:45.91	&	$-$62:04:49.4	&		25	\\
G311.5671+00.3189	&	14:04:22.27	&	$-$61:19:26.7	&		26	\\
G311.9799$-$00.9527	&	14:10:51.43	&	$-$62:25:16.3	&		20	\\
G313.5769+00.3267	&	14:20:08.35	&	$-$60:41:55.3	&		23	\\
G319.8366$-$00.1963	&	15:06:54.88	&	$-$58:32:57.4	&		23	\\
G320.2878$-$00.3069	&	15:10:19.60	&	$-$58:25:07.3	&		25	\\
G320.3767$-$01.9727	&	15:17:39.02	&	$-$59:47:49.2	&		23	\\
G325.3566$-$00.0422	&	15:40:11.95	&	$-$55:23:19.3	&		20	\\
G327.9205+00.0921	&	15:53:32.92	&	$-$53:42:05.4	&		21	\\
G328.2658+00.5316	&	15:53:27.93	&	$-$53:08:33.7	&		21	\\
G333.3752$-$00.2015	&	16:21:05.95	&	$-$50:15:14.0	&		21	\\

\hline
\end{tabular}\\
Notes: Only a small portion of the data is provided here, the full table is only  available in electronic form at the CDS via anonymous ftp to cdsarc.u-strasbg.fr (130.79.125.5) or via http://cdsweb.u-strasbg.fr/cgi-bin/qcat?J/A+A/.

\end{minipage}
\end{center}
\end{table}

\subsection{Observations and data reduction}
\label{sect:observations}

The observations were made using the Mopra 22m Radiotelescope in April and September 2008 towards 499 massive star forming regions. Mopra is located near Coonabarabran, New South Wales, Australia.\footnote{Mopra is operated by the Australia Telescope National Facility, CSIRO.} The telescope is situated at an elevation of 866 metres above sea level, and at a latitude of 31 degrees south. 

The telescope is equipped with a 12 mm receiver with a frequency range of 16 to 27.5 GHz. The UNSW Mopra spectrometer (MOPS) is made up of four 2.2~GHz bands which overlap slightly to provide a total of 8~GHz continuous bandwidth. Up to four zoom windows can be placed within each 2.2~GHz band allowing up to 16 spectral lines to be observed simultaneously. Each zoom window provides a bandwidth of 137~MHz with 4096 channels, which at 12~mm corresponds to a total velocity range of $\sim$2000~\kms\ with a velocity resolution of $\sim$0.4~\kms\ per channel. 

The 8 GHz bandpass was centred at 23.5~GHz providing complete coverage over the 19.5--27.5~GHz range with a corresponding resolution of 2.7--1.8\arcmin\ respectively. Individual zoom windows were deployed to cover: the water maser at 22.235~GHz, the NH$_3$ (1, 1) and (2, 2) inversion lines at 23.694 and 23.722~GHz, two more covering the  NH$_3$ (3, 3) and (4, 4) lines located at 23.870 and 24.139~GHz respectively. Additional zoom windows were placed to cover the thermal methanol lines around 24.9~GHz, and at 23.121~GHz to cover the rare methanol maser transition (Cragg et al. 2004). The remaining windows were used to search for the more common 19.97~GHz methanol maser line (Ellingsen et al. 2004) and HC$_5$N (7--6)  transition at 18.64~GHz. In this paper we will concentrate on the detection statistics of water masers and postpone the analysis of the other transitions to a subsequent publication.

\begin{figure}
\begin{center}
\includegraphics[width=0.45\textwidth]{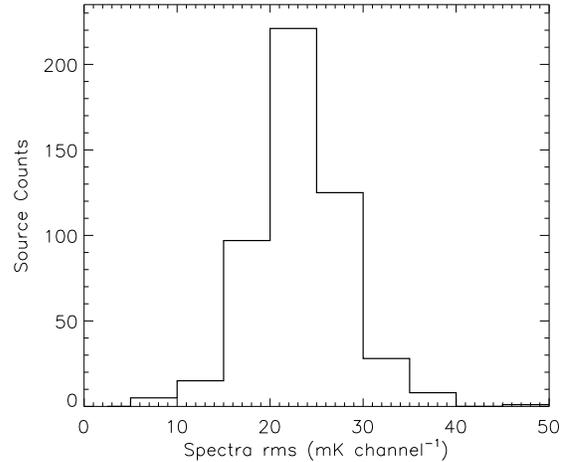}
\caption{\label{fig:field_rms_hist} Histogram of the r.m.s. noise level in the different observations. The bin size is 5 mK channel$^{-1}$.}

\end{center}
\end{figure}

The observations were conducted in position-switched mode towards 499 RMS sources (see Table~\ref{tbl:source_positions} for positions). Each source was observed for a total of ten minutes of on-source integration, split into a number of separate scans consisting of 1 minute on- and 1 minute off-source.  Reference positions were offset from the MSX positions by 1 degree in a direction perpendicular to the Galactic plane. Given that the size of the beam at 12~mm ($\sim$2\arcmin) is much larger than the possible deviations from the global pointing model no pointing checks were performed. System temperatures were between 65-100~K depending on weather conditions and telescope elevation, resulting in typical rms values of $\sim$25~mK per channel (see Fig.~\ref{fig:field_rms_hist} for distribution). 

Cross scans of Jupiter were made to determine the telescope main beam efficiency during our observations which was found to be $\eta_{\rm{MB}}$ $\sim$0.55 at 23~GHz. The corrected antenna temperatures (T$_A^*$) were put on the main beam temperature scale by dividing by the telescope main beam efficiency ($\eta_{\rm{MB}}$). Finally, the main beam temperatures were converted to flux density using a conversion factor of 6.41~Jy K$^{-1}$ (Eqn~7.19;  \citealt{rohlfs2004}). Taking the conversion from T$_A^*$ to Jy into account the nominal 4$\sigma$ detection limit is $\sim$1~Jy per channel. Hereafter all quoted intensities will be given in Janskys. 

\begin{figure*}
\begin{center}

\includegraphics[width=0.45\linewidth]{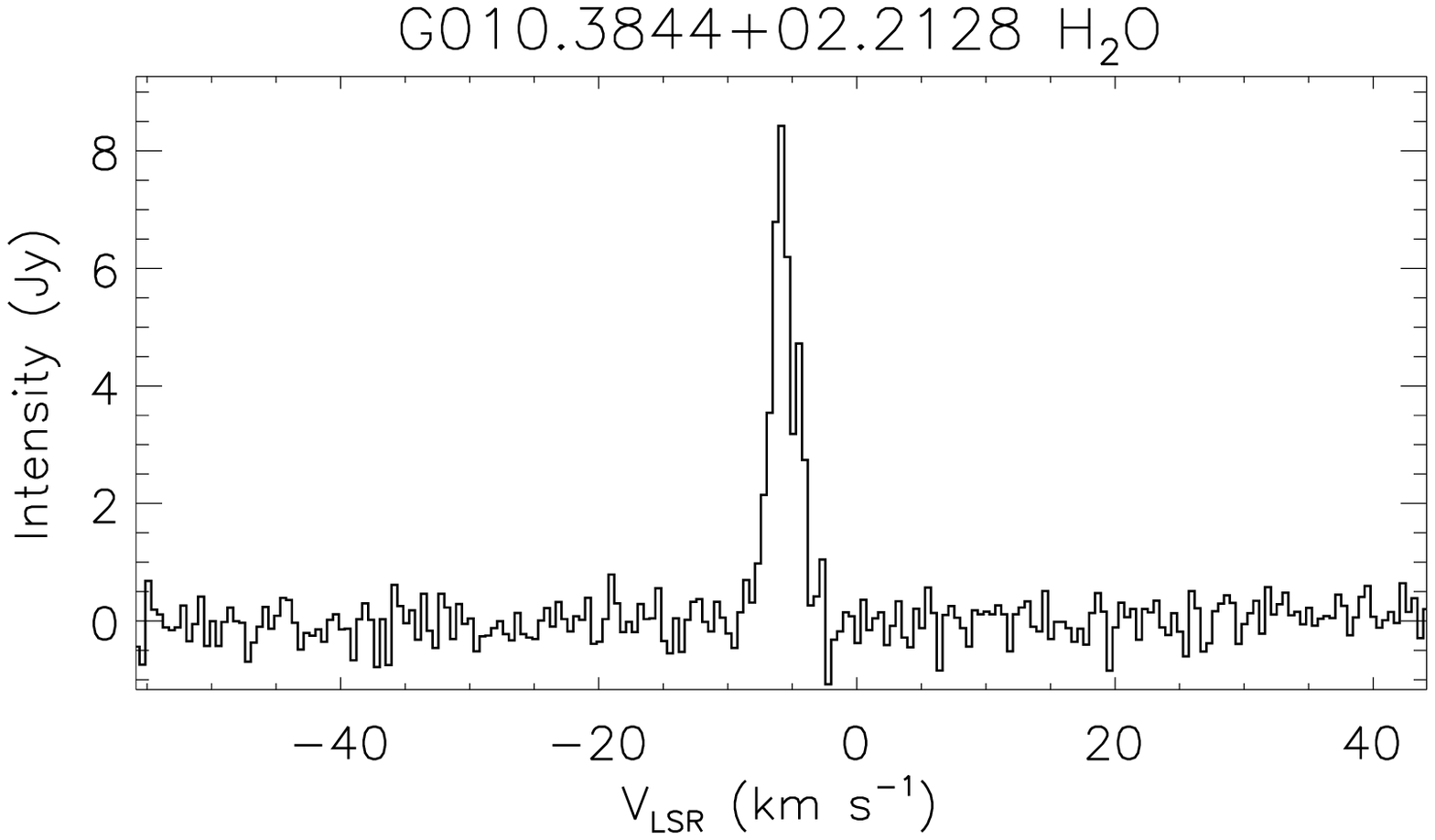} 
\includegraphics[width=0.45\linewidth]{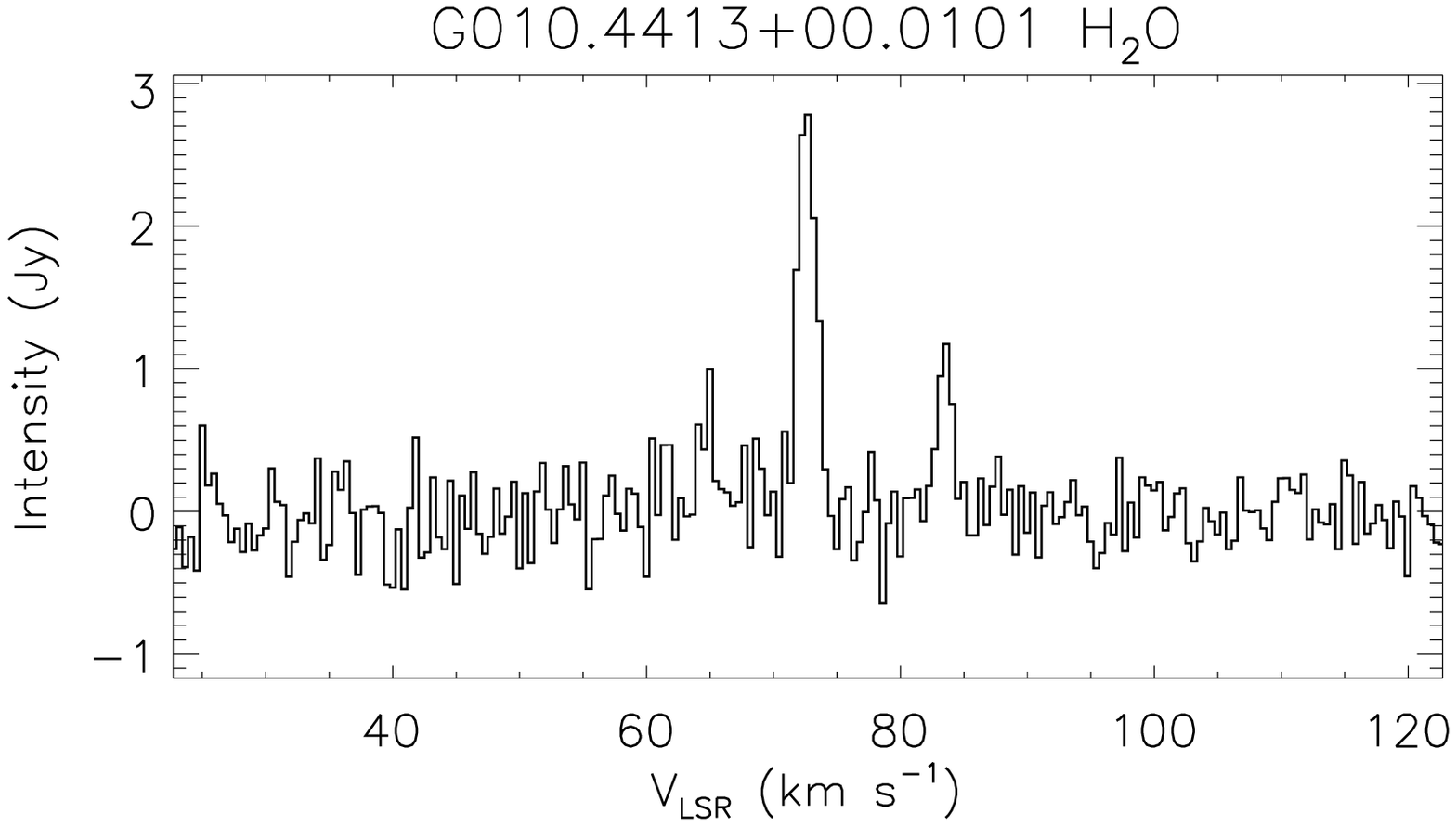}\\  
\includegraphics[width=0.45\linewidth]{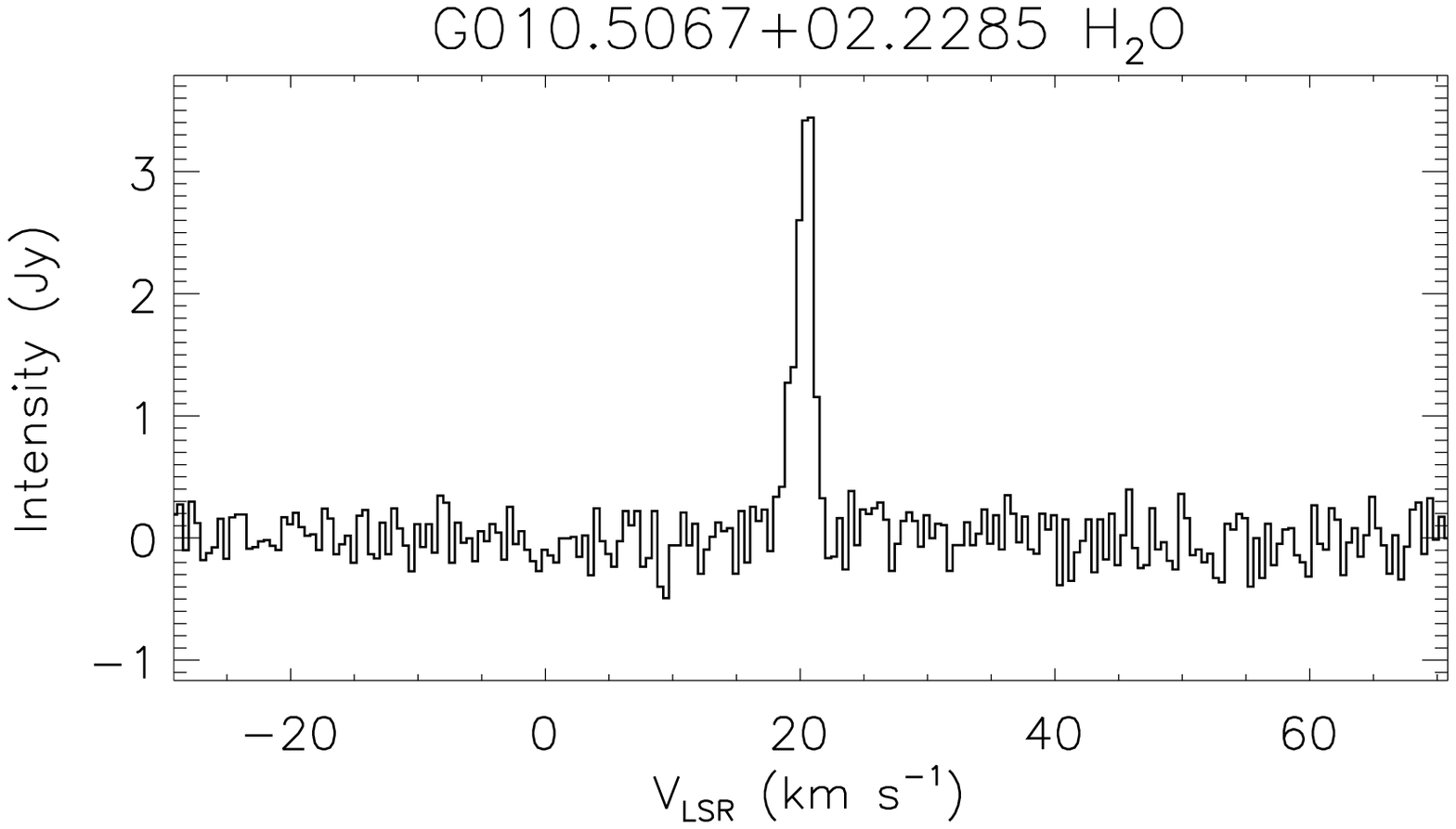} 
\includegraphics[width=0.45\linewidth]{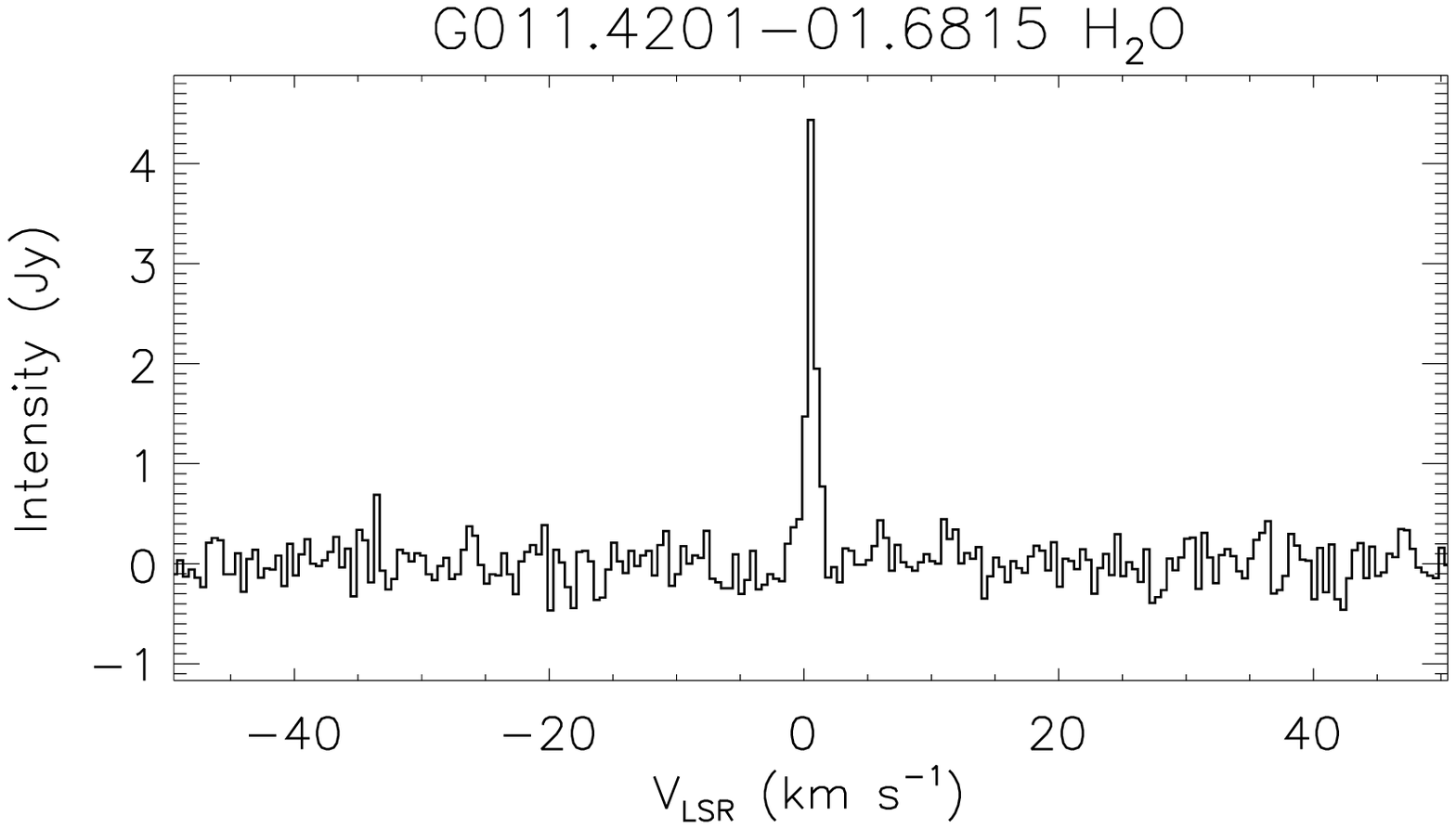}\\  
\includegraphics[width=0.45\linewidth]{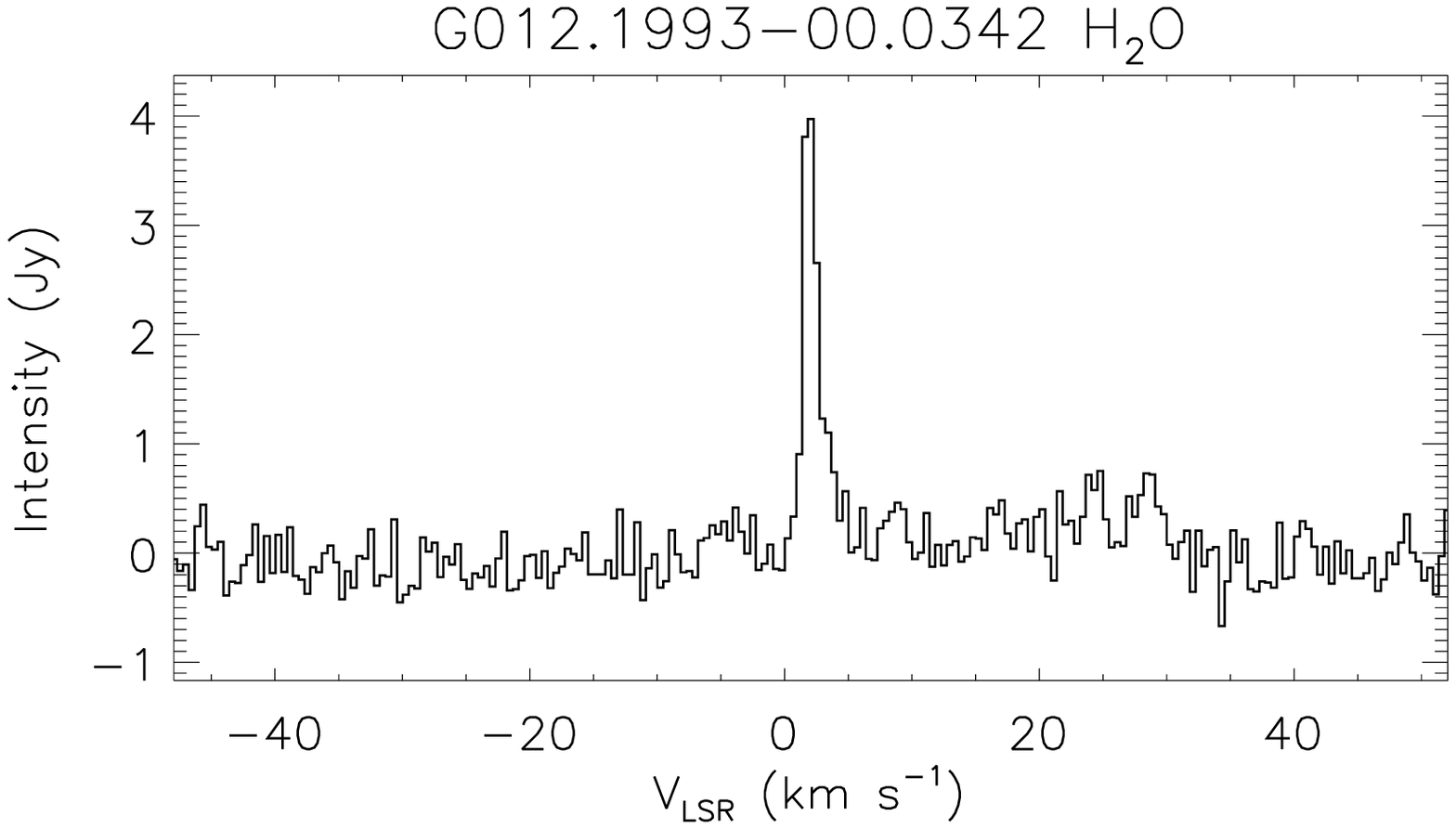} 
\includegraphics[width=0.45\linewidth]{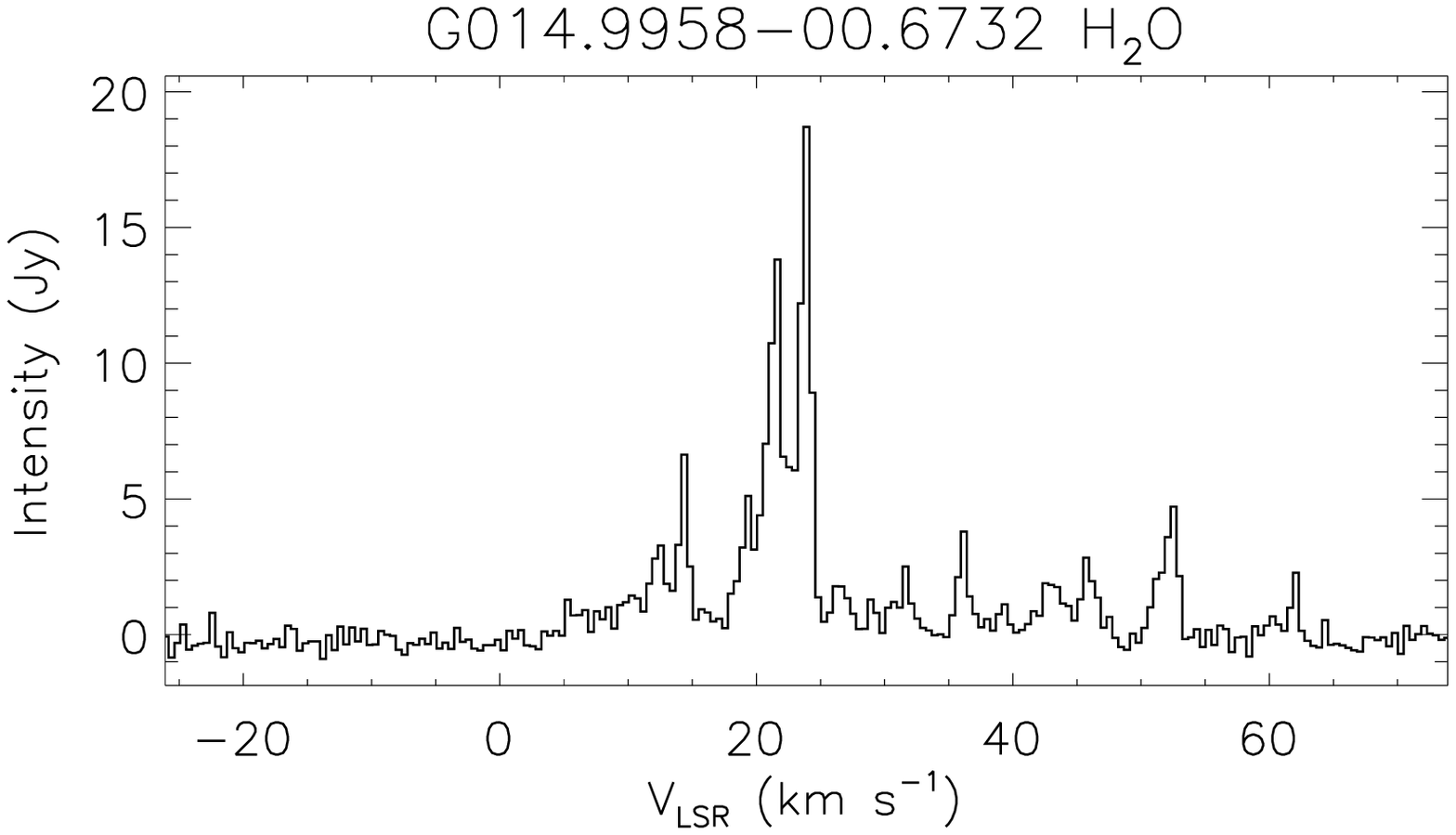} \\ 

\caption{\label{fig:maser_sample} Spectra of detected H$_2$O masers. Only a small portion of the plots are provided here, the full figure is only available in electronic form at the CDS via anonymous ftp to cdsarc.u-strasbg.fr (130.79.125.5) or via http://cdsweb.u-strasbg.fr/cgi-bin/qcat?J/A+A/. 
}
\end{center}
\end{figure*}

The data were reduced using the ATNF spectral line reducing software (ASAP). The reduction steps consisted of  dividing the individual on-off scans to remove sky emission (these scans were then inspected and poor scans were removed), fitting a low-order polynomial to the baseline. Finally, the scans were averaged together to produce a single spectrum for each source.

\section{Results and analysis}

\subsection{Detection statistics and general properties}

We detect maser emission towards 153 of the 499 star forming regions observed. In 10 of these cases two distinct groups of maser emission were observed separated in velocity by $>$ 30~\kms. The emission structure and velocity separation makes it likely that the emission seen in these cases is from two distinct sites of maser emission along the same line of sight, within the $\sim$2
arcmin beam. We have therefore classified these as separate detections and indicate these in the plot presented in Fig.~\ref{fig:maser_sample} and Table~\ref{tbl:maser_parameters} by appending an `\_$n$' to the MSX name (where $n$ = 1 or 2). In all of these cases at least one of the line of sight components is found at a similar velocity as a molecular tracer and is therefore likely to be associated with an RMS source (see Sect.~4.1 for details). However, the origin of these other masers detected along each line of sight is unclear since the majority (8) have velocities outside the velocity range usually associated with molecular gas and so were not covered by our CO observations (see discussion in Sect.~4.3). Only two of these masers are found within the velocity range of our CO observations. One of which is found to have a similar velocity as a weak CO component, whilst no CO emission is seen around the velocity of the other, however, given the difference in resolution this is perhaps not surprising  (CO observations have a resolution of 34\arcsec;~\citealt{urquhart_13co_south}). 

\begin{table*}
\begin{center}
\caption{Parameters of detected masers.}
\label{tbl:maser_parameters}
\begin{minipage}{\linewidth}
\begin{tabular}{lcc....}
\hline
\hline
MSX Name	&	RA	&	Dec	&	\multicolumn{1}{c}{Peak Flux}	&	\multicolumn{1}{c}{\vlsr}	&	\multicolumn{1}{c}{V$_{\rm{min}}$}	&	\multicolumn{1}{c}{V$_{\rm{max}}$}	\\
	&	(J2000)	&	(J2000)	&	\multicolumn{1}{c}{(Jy)}	&	\multicolumn{1}{c}{(\kms)}	&	\multicolumn{1}{c}{(\kms)}	&	\multicolumn{1}{c}{(\kms)}	\\
\hline

G010.3844+02.2128	&	18:00:22.68	&	$-$18:52:08.0	&	8.4	&	-5.8	&	-9.1	&	-2.5\\
G010.4413+00.0101	&	18:08:38.23	&	$-$19:53:57.4	&	2.8	&	72.7	&	63.5	&	84.9\\
G010.5067+02.2285	&	18:00:34.58	&	$-$18:45:17.6	&	3.4	&	20.8	&	17.4	&	22.1\\
G011.4201$-$01.6815	&	18:16:56.86	&	$-$19:51:07.2	&	4.4	&	0.5	&	-1.4	&	1.6	\\
G012.1993$-$00.0342	&	18:12:23.66	&	$-$18:22:51.6	&	4.0	&	2.1	&	-0.2	&	30.0	\\
G013.0105$-$00.1267	&	18:14:22.25	&	$-$17:42:47.8	&	2.4	&	5.3	&	-7.4	&	16.4	\\
G014.9958$-$00.6732	&	18:20:19.43	&	$-$16:13:31.0	&	18.9	&	23.9	&	3.6	&	63.6	\\
G015.0939+00.1913	&	18:17:20.86	&	$-$15:43:47.2	&	8.1	&	30.2	&	23.4	&	33.8\\
G015.1288$-$00.6717\_1	&	18:20:34.75	&	$-$16:06:26.2	&	1.7	&	-41.9	&	-43.0	&	-41.3\\
G015.1288$-$00.6717\_2	&	18:20:34.75	&	$-$16:06:26.2	&	0.7	&	22.2	&	19.7	&	25.6\\
G016.8055+00.8149	&	18:18:25.96	&	$-$13:55:37.9	&	5.1	&	70.4	&	68.0	&	72.5\\
G016.9270+00.9599	&	18:18:08.59	&	$-$13:45:05.7	&	1.7	&	-65.8	&	-68.0	&	-59.1\\
G017.9789+00.2335	&	18:22:49.08	&	$-$13:09:59.0	&	4.7	&	18.1	&	15.8	&	20.3\\
G018.1409$-$00.3021	&	18:25:04.46	&	$-$13:16:26.7	&	0.7	&	13.0	&	10.7	&	13.6\\
G024.2138$-$00.0439	&	18:35:35.80	&	$-$07:46:22.4	&	4.0	&	55.7	&	53.2	&	56.8\\
G025.4018+00.0198	&	18:37:34.19	&	$-$06:41:18.6	&	7.9	&	-13.3	&	-15.9	&	-6.5\\
G026.1094$-$00.0944	&	18:39:17.14	&	$-$06:06:43.9	&	16.4	&	25.1	&	20.6&	28.6\\
G028.3271+00.1617	&	18:42:26.90	&	$-$04:01:23.8	&	1.9	&	31.3	&	29.1	&	32.8\\
G030.4117$-$00.2277	&	18:47:39.04	&	$-$02:20:47.7	&	2.8	&	105.9	&	97.2	&	107.0\\

G030.8667+00.1141	&	18:47:15.81	&	$-$01:47:08.8	&	2.9	&	37.8	&	36.3	&	48.4\\
G188.8120+01.0686	&	06:09:17.85	&	+21:50:49.9	&	10.4	&	-5.9	&	-8.3&	-3.8\\
G200.0789$-$01.6323	&	06:21:47.85	&	+10:39:20.8	&	2.0	&	35.2	&	33.8	&	42.0\\
G224.3494$-$02.0143	&	07:05:12.43	&	$-$11:04:31.0	&	3.8	&	16.4	&	14.6	&	21.4\\
G224.6573$-$02.4978	&	07:04:01.65	&	$-$11:34:12.3	&	0.6	&	11.5	&	9.8	&	13.2\\
G226.2728$-$00.4633	&	07:14:26.80	&	$-$12:03:58.3	&	2.9	&	17.3	&	15.6	&	17.9\\
G229.5711+00.1525	&	07:23:02.08	&	$-$14:41:32.2	&	9.0	&	60.8	&	47.7	&	62.4\\
G234.5093$-$00.3181	&	07:31:08.76	&	$-$19:15:29.8	&	1.6	&	43.5	&	42.0	&	45.4\\
G236.8158+01.9821	&	07:44:27.93	&	$-$20:08:33.0	&	15.2	&	54.4	&	39.4&	64.3\\
G240.3156+00.0715	&	07:44:52.00	&	$-$24:07:41.1	&	18.3	&	65.3	&	62.5&	78.9\\

\hline
\end{tabular}\\

\end{minipage}
\end{center}
Notes: Only a small portion of the data is provided here, the full table is only  available in electronic form at the CDS via anonymous ftp to cdsarc.u-strasbg.fr (130.79.125.5) or via http://cdsweb.u-strasbg.fr/cgi-bin/qcat?J/A+A/.

\end{table*}

Taking these cases into account, we have detected a total of 163 masers above a 4$\sigma$ noise level ($\sim$1~Jy), which corresponds to an overall detection rate of $\sim$32\%. We conducted a literature search using SIMBAD and found that approximately 75\% of these masers were previously unknown. In Fig.~\ref{fig:maser_sample} we present plots of all the detected masers. Where two masers have been detected along the same line of sight we present a separate plot for each. To avoid confusion the velocity range for a number of these multiple detections has been reduced. As can be seen from the spectra presented in this figure, the maser emission generally consists of either a single component or a group of emission peaks spread over a clearly identifiable velocity range. 
In order to include all emission components associated with each maser the plots presented in Fig.~\ref{fig:maser_sample} cover a velocity range of $\sim$100~\kms\ and are centred on the velocity of the strongest component.

\begin{figure*}
\begin{center}
\includegraphics[width=0.45\textwidth]{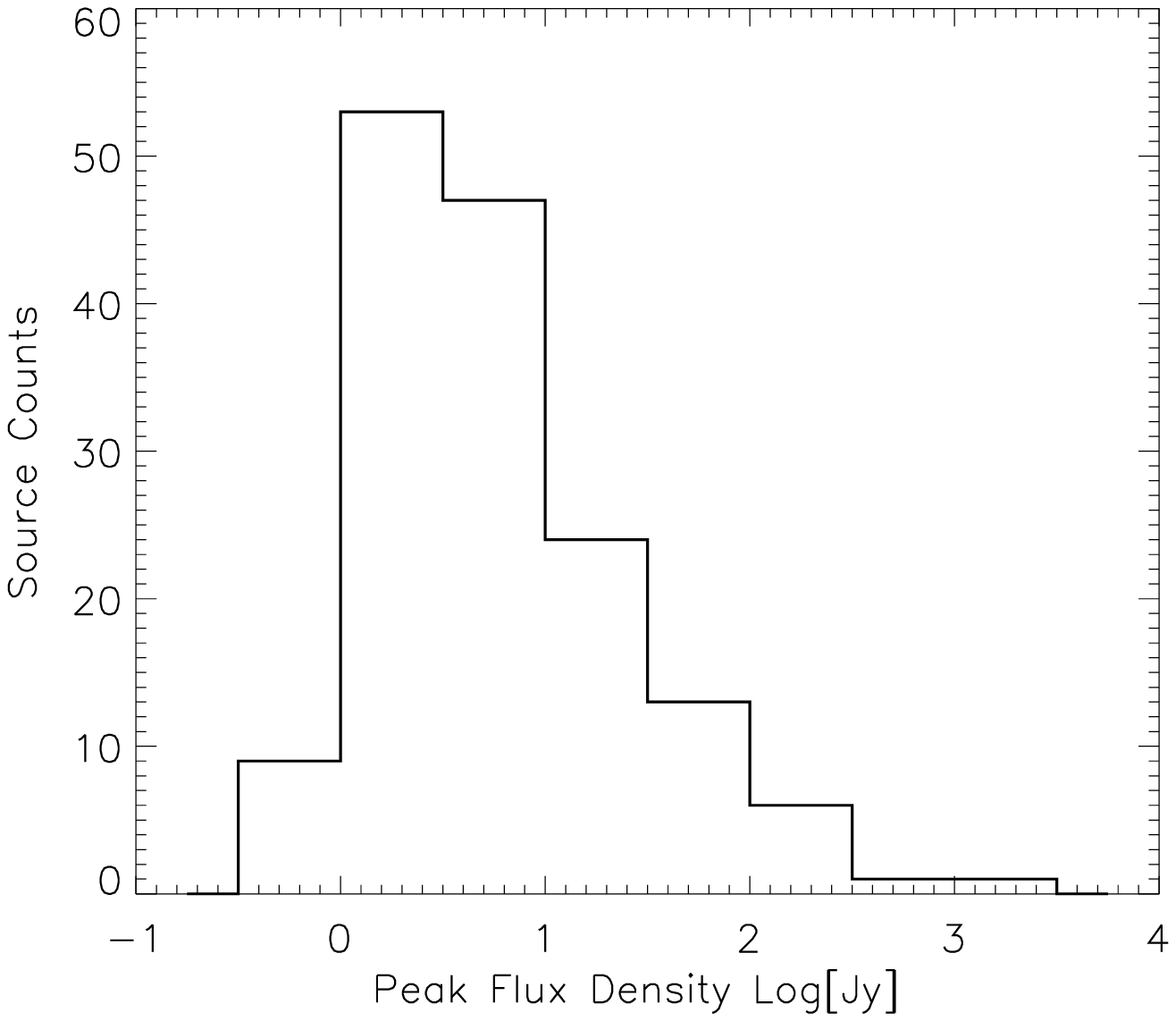}
\includegraphics[width=0.45\textwidth]{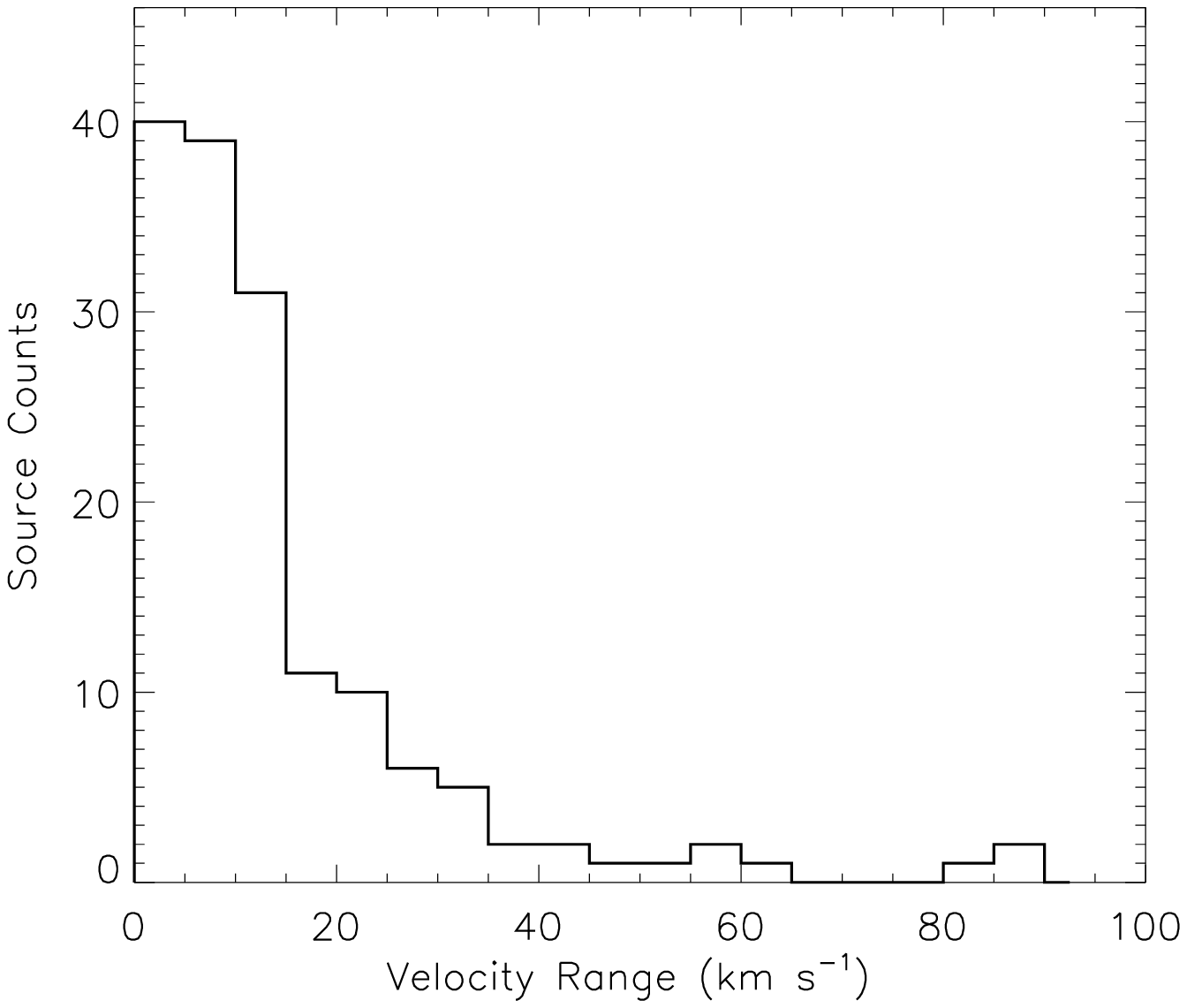}
\caption{\label{fig:velocity_range_hist} Histograms of the  H$_2$O maser parameters. In the left panel we present the histogram of the peak intensity distribution (the bin size is 0.5 dex) and in the right panel we present a histogram of the H$_2$O maser groups dispersions (binned using a value of 5 km s$^{-1}$).}

\end{center}
\end{figure*}

In Fig.~\ref{fig:velocity_range_hist} we present histograms of the distribution of the peak flux density and maser group velocity dispersion (defined as a number of discrete components within 30 \kms\ of each other). The distribution of peak flux densities spans a range of 4 orders of magnitude from $\sim$0.5~Jy up to a couple of thousand Jy, however, more typical values are between 1--10~Jy. The histogram of the distribution of maser velocity dispersion, shown in the right panel of Fig.~\ref{fig:velocity_range_hist}, illustrates the large spread in velocity of maser emission. Maser groups consisting of a single component have an intrinsic width of as little as a few \kms\ with multiple component groups having velocity spreads of up to 90~\kms. However, these are the extremes with the majority of water masers having velocity ranges of 10--15~\kms. This large velocity dispersion is rather typical for water masers and is one of the reasons a connection with molecular outflows is often inferred (e.g., \citealt{elitzur1989,menten1996}).

In Table~\ref{tbl:maser_parameters} we present the parameters of the detected H$_2$O masers; we give the MSX name and J2000 co-ordinates in Columns 1--3. In Column 4 and 5 we give the peak flux density and the velocity of the peak emission. We give the minimum and maximum velocity range of the emission in Columns~6 and~7.

\section{Discussion}

\subsection{RMS-maser associations}

We have detected and identified 163 water masers towards RMS sources. Not all will necessarily be associated, however, as some are likely to be  chance alignments along the line of sight. Given the spatial resolution of these observations, we are unable to confirm associations but we can identify likely RMS-H$_2$O associations by considering the velocities of the maser and the RMS sources (determined from CO observations, see \citealt{urquhart_13co_south,urquhart_13co_north} for details) and look for a velocity correlation. 

\begin{figure}
\begin{center}
\includegraphics[width=0.45\textwidth]{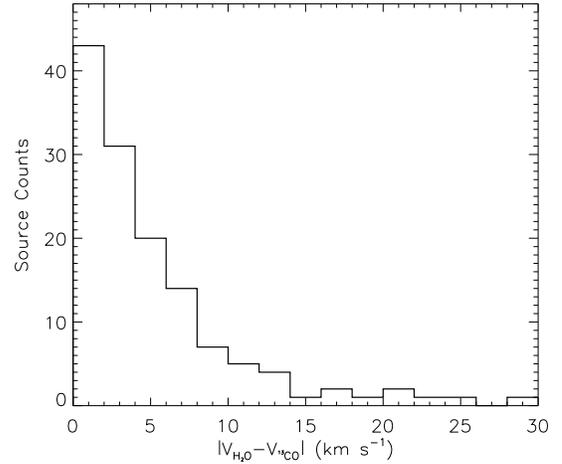}
\caption{\label{fig:delta_vlsr_hist} Histogram of the differences between the velocity of the detected H$_2$O maser and the  $^{13}$CO velocity of the RMS source within the beam. The bin size is 2 km s$^{-1}$.}

\end{center}
\end{figure}

In an effort to identify likely RMS-H$_2$O maser associations we compared the peak maser velocity with the velocity of the RMS source along each line of sight. The distribution shown in Fig.~\ref{fig:delta_vlsr_hist} illustrates the excellent correlation between the velocity of the peak intensity component in each maser group and the velocity of the molecular cloud within which the RMS source resides. The distribution is strongly peaked at $\leq2$~\kms, and falls off steadily until reaching a background level at $\sim$14~\kms. The shape of the histogram corresponds to that of a one-sided Gaussian and we have therefore determined the standard deviation from a Gaussian fit to the profile.

We consider an RMS source and H$_2$O maser to be associated if the difference in their velocities is less than 3$\sigma$, where $\sigma$ is the standard deviation of the distribution  $\sim$5~\kms\ (cf. a median difference of 4.5~\kms;~\citealt{kurtz2005}). Applying this criterion (i.e., $\Delta$\vlsr\ $<$ 15~\kms) we find 131 RMS-H$_2$O maser associations. No velocity has been assigned to five RMS sources due to the presence of multiple CO components in the spectra, which makes it difficult to determine a unique velocity (\citealt{urquhart_13co_south}). However, it is still possible that the masers detected towards these sources are associated. To investigate these possible associations we compared the velocities of the peak maser components to the velocities of the detected CO components seen towards all five RMS sources. We found the peak component velocity of the maser within 15~\kms\ of at least two of the CO peaks present in four of the five cases, and therefore consider these to be possible associations. Unfortunately, we were not able to use the presence of the masers to identify the CO component associated with the RMS source in these cases. Including the 131 RMS sources already found to have an associated water maser this brings the total to 135. The remaining 27 masers are likely to be associated with other star forming regions that happen to be located along the same line of sight as the RMS source.

\subsection{Comparison of RMS-H$_2$O detection rate}

Our detection rate for finding water masers associated with RMS sources is $\sim$$27\pm 2$\%. Comparing this value with other detection rates reported from previous H$_2$O surveys we find it is towards the lower end of a range of values (i.e., 20--67\%). For example, observations by \citet{churchwell1990} and \citet{kurtz2005} towards bright IRAS sources reported relatively high detection rates of 67\% and 55\% respectively, while \citet{palla1991} and \citet{codella1995} reported detection rates of only 26\% and 20\% respectively.

All of these surveys where made towards IRAS sources that satisfied the \citet{wood1989} colour selection criteria for identification of UCHII regions (i.e., ${\rm{log}}(F_{60}/F_{12}) \geq 1.30$ and ${\rm{log}}(F_{25}/F_{12}) \geq 0.57$). However, most of the \citet{churchwell1990} sources, and all of the \citet{kurtz2005} sources,  have IRAS 100~\mum\ $>$ 1000~Jy and IRAS 60~\mum\ $>$ 100~Jy and are therefore much more biased towards brighter sources than the \citet{palla1991} and \citet{codella1995} surveys. Additionally, the \citet{churchwell1990} and \citet{kurtz2005} surveys are approximately 10 times more sensitive than the other two surveys. If a similar sensitivity  to that of Palla and Codella were applied to the \citet{churchwell1990} and \citet{kurtz2005} surveys, their detection rates would be reduced to 35\% and 36\%, respectively (see \citealt{kurtz2005} for more details). Given the differences in sensitivity, source selection of these surveys, and statistical noise we find our results are in good agreement with all of the surveys mentioned.

While \citet{churchwell1990} and \citet{kurtz2005} concentrated on the UCHII  stage (identified from their radio continuum emission), the \citet{palla1991} and the \citet{codella1995} samples are likely to contain a mixture of nearby low-mass YSOs and more distant HII regions and massive YSOs. Having a well selected sample of bona fide YSOs and UCHII regions associated with water masers we can now go a step further and look at how the association rates vary between the different evolutionary stages. In Table~\ref{tbl:source_types} we present a breakdown of the number of RMS-H$_2$O maser associations as a function of source classification (the errors are calculated assuming a binomial distribution). We can immediately see that the detection rates of H$_2$O masers for all UCHII regions and YSOs are very similar. This would suggest that the conditions necessary for maser activity are equally likely in these two stages of the star formation process. The lower detection rate seen towards the Young/old class of sources is probably a reflection of the number of evolved stars within this group as discussed in Sect.~\ref{sect:source_selection}, however, given the small number of statistics this may not be all that significant. 

\begin{table}
\begin{center}
\caption{Summary of number of sources observed, maser detections and detections rates by source classification.}
\label{tbl:source_types}
\begin{minipage}{\linewidth}
\begin{tabular}{lccc}
\hline
\hline
Source type	& \# of Obs. & \# of Association	& Detection rate \\

\hline

YSOs		&	233	&	60	&	$\sim$$25\pm 4$\%		\\
HII/YSOs	&	105	&	34	&		$\sim$$32\pm 5$\%	\\
Young/Old	&	40	&	7	&	$\sim$$18\pm 7 $\%	\\
HII regions	&	125	&	34	&	$\sim$$27\pm 4 $\%	\\

\hline
\end{tabular}\\

\end{minipage}
\end{center}
\end{table}

It is somewhat surprising to find the detection rate for UCHII regions and YSOs is effectively the same, given that one might expect the environments to be very different. Water masers are widely thought to be associated with molecular outflows and/or accretion, however, both of these phenomenon are expected to be disrupted and eventually halted altogether once the HII region begins to form. We compared the measured parameters (e.g., maser velocity dispersions, peak intensity) for the UCHII regions and MYSO candidates to try and identify any differences that might give an insight into their nature, however, we found no significant differences - Kolmogorov-Smirnov (KS) tests of these parameters are consistent with the masers being drawn from the same overall population. 

The detection rate found towards YSOs and UCHII regions is double the rate reported by \citet{wang2006} ($\sim$12\%) from observations of 140 compact cores found within a sample of infrared dark clouds (IRDCs). These observations were made with the VLA and had a sensitivity of $\sim$0.1 Jy beam$^{-1}$ channel$^{-1}$, which is a factor of a few times more sensitive than the observations presented here. The rate of detection of water masers is therefore significantly higher for star forming cores than for cores found in IRDCs, which are presumably at an earlier stage in their evolution, and many of which, are likely to be of relatively low luminosity.

\subsection{Galactic distribution}

\begin{figure}
\begin{center}

\includegraphics[width=0.45\textwidth]{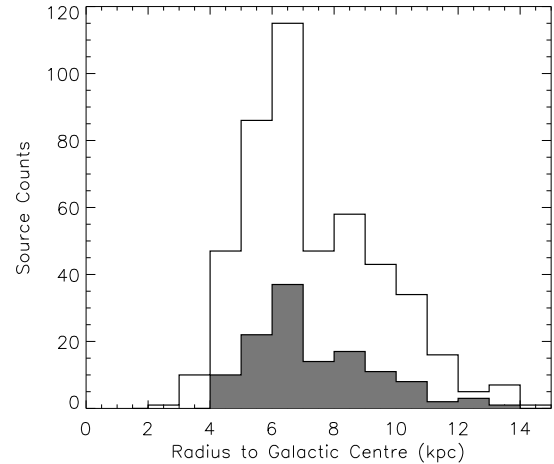}

\caption{\label{fig:rgc_hist} Histograms of the distance from the Galactic centre of all YSOs and HII regions (outlined by solid line) and those associated with H$_2$O masers (filled histogram). The bin size is 1~kpc.}

\end{center}
\end{figure}

\begin{figure}
\begin{center}
\includegraphics[width=0.45\textwidth]{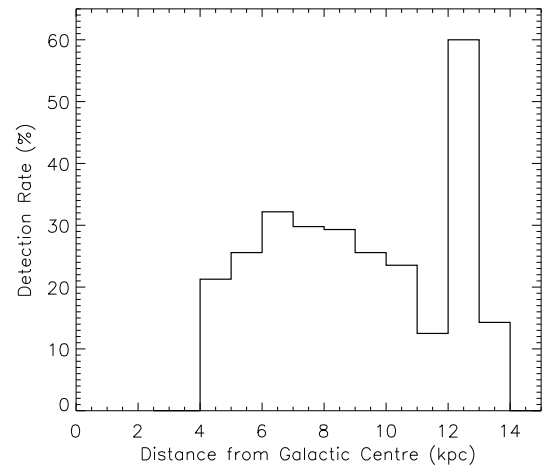}

\caption{\label{fig:detection_hist} Plot of the detection rate of H$_2$O masers as a function of distance from the Galactic centre of all YSOs and HII regions. The bin size is 1~kpc.}

\end{center}
\end{figure}

Comparing the Galactic longitude and latitude distributions we find no significant differences between the RMS sources associated with masers and those without. However, we do find a difference in the proximity to the Galactic centre between the two samples, with a much higher proportion of those associated with H$_2$O masers being located at smaller Galactocentric radii than those that are not associated. In Fig.~\ref{fig:rgc_hist} we present a histogram of the distribution of Galactocentric  radii for the whole sample (outlined by the solid line) and the RMS-H$_2$O maser associated sources (filled histogram). The distributions of the two populations are similar with both showing the same features. The two most interesting features are the two peaks located at $\sim$6--7~kpc and  8--9~kpc. These distributions are very similar to the radial distribution of an IRAS selected sample of UCHII regions reported by \citet{bronfman2000} from CS observations.

The 6--7~kpc peak correlates with a peak in the radial distribution of a sample of southern infrared dark clouds (IRDCs) reported by \citet{jackson2008}. This peak is at a larger radial distance from the Galactic centre than found for a sample of northern IRDCs which peaks at $\sim$5~kpc. The difference galactocentric distributions between the Galactic first and fouth quadrants led \citet{jackson2008} to conclude that these features are more likely to be associated with the Scutum-Centaurus arm in a two arm Galactic model (e.g., \citealt{drimmel2000,drimmel2001}) than part of a molecular ring of material located at $\sim$5~kpc (see \citealt{jackson2006} and references therein). The coincidence of the peak in the distribution of our sample of UCHII regions and MYSO candidates, the sample of IRAS selected UCHII regions (\citealt{bronfman2000}) and IRDCs (\citealt{jackson2008}) with the Scutum-Centaurus arm is consistent with observations of external spiral galaxies where the formation of OB stars is seen to be exclusively associated with the spiral arms. 

The second peak located $\sim$8--9~kpc is most likely due to a high number of sources found in the solar neighbourhood ($\simeq$ 0.5~kpc from the Sun). The distribution of the RMS sources associated with H$_2$O masers shares some similarities with the distribution of the general RMS population, however, comparing it with the distribution of unassociated RMS sources reveals them to be significantly different. A KS test of these samples results in only a 6\% probability that they are drawn from the same population.

Looking in a little more detail we find that the detection rate varies as a function of distance from the Galactic centre. This is illustrated in Fig.~\ref{fig:detection_hist} which shows the detection rate to be approximately constant ($\sim$27\%)  between 4--6~kpc, after which it increases to $\sim$37\% between 6--7~kpc, before declining steadily to $\sim$24\% at 11~kpc. After 11~kpc the detection rate falls off to $\sim$15\%, however, the low number statistics for distances greater than this result in the spike between 12 and 13~kpc. A long term study of water masers towards YSOs by \citet{brand2003} found the detection rate, and maser stability, was significantly higher for more luminous young stars ($\geq 3 \times 10^4$ \lsun) than for lower luminosity YSOs. The increase in detection rates for our sample of young massive stars located within the solar circle probably  reflects the presence of a higher proportion of more luminous sources. Since the majority of our sources are located within the solar circle, and are thus subject to the distance ambiguity problem, reliable distances are not yet available for many sources. We are therefore not currently in a position to investigate this inference in more detail.

\begin{figure*}[!ht]
\includegraphics[height=0.99\textwidth,angle=270, trim=0 10 0 30]{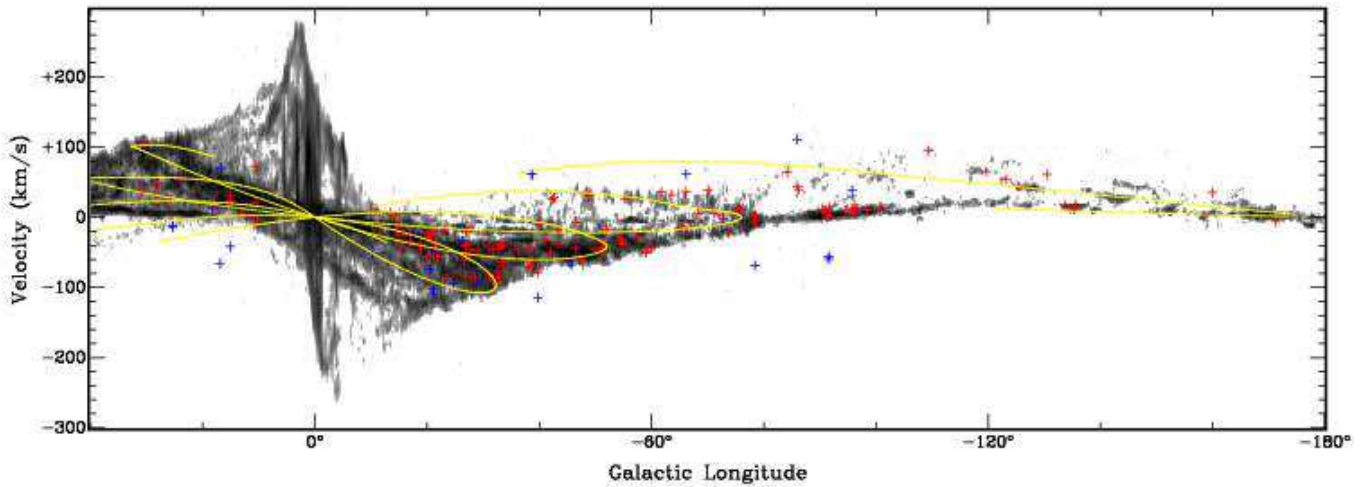}

\caption{Galactic longitude-velocity plot showing the velocities of all H$_2$O masers detected as a function of Galactic longitude. Masers associated with an RMS source are shown in red, while unassociated H$_2$O masers are shown in blue. The distribution of molecular material is shown in grey scale (\citealt{dame2001}) for comparison. The location of the spiral arms taken from the model by \citet{tayor1993} and updated by \citet{cordes2004} are over-plotted in yellow.}

\label{fig:h2o_distribution}

\end{figure*}

In Fig.~\ref{fig:h2o_distribution} we present a plot of the distribution of $^{12}$CO as a function of Galactic longitude and V$_{\rm{LSR}}$ (\citealt{dame2001}). We have over-plotted the velocities of all detected masers; we distinguish between masers associated with an RMS sources and the unassociated masers by red and blue crosses respectively. The image presented in Fig.~\ref{fig:h2o_distribution} illustrate the excellent correlation of RMS-H$_2$O maser associations with the molecular material in the Galaxy  as traced by the integrated $^{12}$CO emission. There is also a reasonable level of correlation between the maser velocities and the spiral arm velocities as modelled by \citet{tayor1993} and \citet{cordes2004}.

In total 27 H$_2$O masers were not associated with an RMS source (i.e., the velocity of their peak component is outside than 15 \kms\ criteria required for an association), this corresponds to $\sim$17\% of all the detected masers. These unassociated masers are the result of chance alignments along the line of sight. The high number of unassociated masers is due to the low resolution of these observations and the density of star forming regions located in the inner Galaxy. These masers fall into two catagories; those with velocities within the velocity range covered by the molecular line observations (see \citealt{urquhart_13co_south,urquhart_13co_north} for details), and those at velocities outside the range normally associated with molecular gas in the Galaxy. We find weak CO emission at similar velocities for a few of the masers that have velocities within the Galactic range, however, for the majority no such emission has been detected. This is probably due to the difference in resolution of the CO observations and the maser observations presented in this paper  (which are $\sim$34\arcsec\ and $\sim$2\arcmin\ respectively), with the star forming regions associated with the maser lying outside the CO beam. 

The second category include masers at velocities outside those typically associated with the Galactic molecular gas. These can be clearly seen in  Fig.~\ref{fig:h2o_distribution} to have significantly different velocities  (e.g., $l$=10.4\degr\ and \vlsr=72.7~\kms\ and $l$=16.8\degr\ and \vlsr=70.4~\kms). Since the conditions required to excite water maser emission are high densities ($\geq 10^8~{\rm{cm}}^{-3}$) and temperatures up to ~400~K the lack of any large scale molecular gas is puzzling. One possibility is that these masers are associated with evolved stars, however, the nature of these masers remains uncertain and further observations will be required to ascertain their origin.

\section{Summary and conclusions}
\label{sect:summary}

We have presented the result of a set of water maser observations towards a sample of 499 massive star forming regions. We detected a total of 163 masers above a 4$\sigma$ sensitivity limit of $\sim$1~Jy, which makes this one of the most sensitive water masers surveys so far conducted. The majority of masers detected were previously unknown ($\sim$75\%). We present plots for all detected H$_2$O masers and tabulate their properties.

By comparing the peak intensity velocities of the detected H$_2$O masers with the velocities of the RMS sources we have identified 135 RMS-H$_2$O maser associations. The overall detection rate towards our sample of young massive star forming regions is $\sim$27\%. This is in agreement with other water maser surveys taking into account their various selection criteria and sensitivities. We find similar detection rates for MYSO candidates and UCHII regions suggesting that maser activity is equally possible during these two stages of the star formation process. The detection rate as a function of distance from the Galactic centre is significantly enhanced within the solar circle, peaking at $\sim$37\% $\sim$ 6~kpc, possibly indicating the presence of a high proportion of more luminous YSOs and HII regions.  Comparing our results with those reported by previous surveys towards massive star forming regions we find them to be consistent. 

These observations are a first step in our programme of follow-up observations designed to examine the global characteristics of this galaxy-wide sample of massive young stars. Further observations are planned to follow up the H$_2$O masers detected from the observations presented in this paper the results of which will be the focus of a subsequent paper.

\begin{acknowledgements}

The authors would like to thank the Director and staff of the Paul Wild Observatory for their
assistance during the preparation of our observations. We would like to thank the referee for some very useful comments and suggestions. JSU is supported by a STFC PDRA and CSIRO OCS fellowship. This research would not have been possible without the SIMBAD astronomical database service operated at CDS, Strasbourg, France and the NASA Astrophysics Data System Bibliographic Services.
\end{acknowledgements}

\bibliography{12608.bib}

\bibliographystyle{aa-package/bibtex/aa}

\end{document}